\documentstyle[referee]{mn}

\begin{document}

%

\title{The asymptotic collapsed fraction in an eternal universe}

\author[]{Hugo Martel$^1$ and Paul R. Shapiro$^1$\\
          $^1$ Department of Astronomy, University of Texas, Austin, 
          TX 78712, USA\\ E-mail: hugo@simplicio.as.utexas.edu (HM);
          shapiro@astro.as.utexas.edu (PRS)}

\maketitle

\begin{abstract}
We calculate the maximum fraction
of matter which is able to condense out of the expanding background
universe by gravitational instability -- the asymptotic collapsed fraction
-- for any universe which is unbound and, hence, will expand forever.
We solve this problem by application of a simple, pressure-free,
spherically symmetric, nonlinear model for the growth of density fluctuations
in the universe. This model
includes general kinds of Friedmann universes, such as the open,
matter-dominated universe and those in which there is an extra, uniform
background component of energy-density (e.g. the cosmological constant
or so-called ``quintessence''), perturbed
by Gaussian random noise matter-density fluctuations. These 
background universes all
have the property that matter-domination eventually gives way either to 
curvature-domination or domination by the positive energy density
of the additional background component. When this happens, gravitational 
instability is suppressed and, with it, so is the growth of the collapsed
fraction. 

Our results serve to identify a limitation of
the well-known Press-Schechter approximation for the time-dependent mass 
function of cosmological structure formation. In the latter approximation,
the mass function determined from the predicted collapse of
positive density fluctuations is multiplied by an ad hoc correction
factor of 2 based upon an assumption that every positive density fluctuation
which is fated to collapse will simultaneously accrete an equal share of
additional matter from nearby regions of compensating negative density
fluctuation. The model presented here
explicitly determines the actual value of the
factor by which any positive density fluctuation which ever collapses will
asymptotically increase its mass by accreting from a compensating
underdensity which surrounds it. We show that,
while the famous factor of 2 adopted
by the Press-Schechter approximation is correct for an Einstein-de~Sitter
universe, it is not correct when
the ``freeze-out'' of fluctuation growth inherent in the more general
class of background universes described above occurs. When
``freeze-out'' occurs, the correction factor reduces to unity and the standard 
Press-Schechter approximation must overestimate the collapsed fraction.

To illustrate this effect, we apply our model to currently viable versions of
the Cold Dark Matter (CDM) model for structure formation, with primordial
density fluctuations in accordance with data on cosmic microwave background anisotropy from the {\sl COBE} satellite DMR experiment. For
$H_0=70\rm\,km\,s^{-1}Mpc^{-1}$ and matter-density parameter $\Omega_0=0.3$,
the open, matter-dominated CDM model and the flat CDM model with nonzero
cosmological constant yield asymptotic collapsed fractions on the galaxy
cluster mass-scale of $10^{15}M_\odot$ and above of 0.0361 and 0.0562, 
respectively, only 55\% of the values
determined by the Press-Schechter approximation. These results have 
implications for the use of the latter approximation to compare the
observed space density of X-ray clusters today with that predicted
by cosmological models.

\end{abstract}

\begin{keywords}
cosmology: theory --- galaxies: clusters: general ---
galaxies: formation --- gravitation --- large-scale structure of the universe
\end{keywords}

%

\def \Omo  {\Omega_0}
\def \Omx  {\Omega_{{\rm X}0}}
\def \Omi  {\Omega_i}
\def \Omxi {\Omega_{{\rm X}i}}

\section{INTRODUCTION}

Explaining the origin and evolution of galaxies and large-scale structure
and determining the fundamental properties of the background universe
are the primary goals of modern cosmology.
The most common assumption is that the structure
we observe today (density structures such as galaxies, clusters, and 
voids, as well as velocity structures such as the Virgocentric Infall or 
that associated with the
Great Attractor), results from the growth, by gravitational
instability, of small-amplitude, primordial density
fluctuations present in the universe at early times. 
These fluctuations are normally assumed to originate from a Gaussian random
process. In this case, they can be described
as a superposition of plane-wave density 
fluctuations with random phases. One important
property of these initial conditions is that overdense and underdense
regions occupy equal volumes (in other words, their filling 
factors are 1/2). Since the density is nearly uniform at early times, 
overdense and underdense regions also contain the same mass. 

The gravitational instability scenario makes the following predictions:
overdense regions, because of their larger gravitational field, will 
decelerate faster than the background universe, resulting in an
increase of their density contrast relative to the background. If this
deceleration is large enough, these regions will turn back and recollapse
on themselves, resulting in the formation of positive density
structures such as galaxies and clusters. The opposite phenomenon occurs in
underdense regions. These regions decelerate more slowly than the background
universe, thus getting more underdense, and eventually become
the cosmic voids we observe today.

In this paper, we investigate the asymptotic collapsed fraction, defined
as the fraction of the matter in the universe that will eventually end up
inside collapsed objects. Obviously, this makes sense only in an 
unbound universe.
Naively, we might think that the asymptotic
collapsed fraction
will be equal to 1/2, since half the matter is located in overdense
regions at early times. This ignores two important effects. First, 
some overdense regions might be unbound, and second, matter located 
inside underdense regions could be accreted by collapsed objects.
The importance of these effects depends upon the particular
background universe in which these structures form. Consider, for instance,
an Einstein-de~Sitter universe. In this case, the background density is
exactly equal to the critical density, and therefore all overdense regions
are bound, and will eventually collapse. Furthermore, it can easily be shown
that any mass element located inside an underdense region is gravitationally
bound to at least one overdense region. Consequently, all the matter inside
underdense regions will eventually be accreted by collapsed objects, and
the asymptotic collapsed fraction is unity. This is not true, however, for a 
background universe with mean density below that of an Einstein-de~Sitter
universe.

Interest in models of the background universe in which the matter density
is less than the critical value for a flat, matter-dominated universe is
now particularly strong, on the basis  of several lines of evidence which
can be reconciled most economically if $\Omo<1$, where $\Omo$ is
the present mean matter density in units of the critical value.
(For reviews and references, see, e.g., Ostriker \& Steinhardt 1995;
Turner 1998; Krauss 1998; Bahcall 1999). 
Arguments in favor of a flat universe
with $\Omo<1$ in which a nonzero cosmological constant makes up the difference 
between the matter density and the critical density have been significantly
strengthened recently by measurements of
the redshifts and distances of Type Ia SNe, which are best 
explained if the universe is expanding at an accelerating rate, consistent
with $\Omo=0.3$ and $\lambda_0=0.7$, where $\lambda_0$ is the vacuum
energy density in units of the critical density at present 
(Garnavich et al. 1998a; Perlmutter et al. 1998).
When combined with measurements of the angular power spectrum of
the cosmic microwave background (CMB) anisotropy, these Type Ia SN results
can be used to restrict further the range of models for the mass-energy content
of the universe. In particular, while the SN data alone are better fit
by a flat model with $\Omo<1$ and a positive cosmological constant
than by an open, matter-dominated model with no cosmological
constant (e.g. Perlmutter et al. 1998), the combined information from Type Ia
SNe and the CMB significantly strengthens the case for a flat model with
cosmological constant over that for an open, matter-dominated model
(e.g. Garnavich et al. 1998b). Exotic alternatives to the well-known 
cosmological constant which might also contribute positively
to the total cosmic energy density and thereby similarly affect the mean
expansion rate have also been discussed, sometimes referred to as
``quintessence'' models (e.g. Turner \& White 1997; Caldwell,
Dave, \& Steinhardt 1998). Such models can also explain the presently
accelerating expansion rate indicated by the Type Ia SNe, while satisfying
several other constraints which suggest that $\Omo<1$. The results
from Type I SNe and CMB anisotropy combined can be used to constrain
the range of equations of state allowed for this other component of energy
density $\rho_x$, with pressure $p_x=w_x\rho_xc^2$. The current results 
favor a flat universe with $\Omo<1$ and an equation
of state for the second component with a value of
$w_x\approx-1$ (where $w_x=-1$ for a cosmological constant) 
favored over larger values of $w_x$ (such as would
describe topological defects like domain walls, strings, or textures),
although the restriction of the range allowed for $w_x$ is not yet very
precise (Garnavich et al. 1998b).

Consider now an unbound universe with a matter density 
parameter $\Omega$ with present value $\Omo<1$. 
In such a universe, the critical density exceeds the
mean density, and therefore some overdense regions are unbound. 
The asymptotic collapsed fraction could still be unity if all the matter in 
overdense, unbound regions plus all the matter in underdense regions
is accreted. This will never be the case, however. 
In such a universe, the density parameter $\Omega$ is near unity at
early times, and structures can grow. Eventually $\Omega$
drops significantly below unity, and a phenomenon known
as ``freeze-out'' occurs. In this regime, density fluctuations do not grow
unless their density is already significantly larger than the background
density. After freeze-out, accretion
by collapsed objects will be very slow, and most of the unaccreted matter
will remain unaccreted. The 
asymptotic collapsed fraction will therefore be less than unity.

The asymptotic collapsed fraction is a quantity which is relevant to
modern attempts to interpret observations of cosmic structure
in at least two ways. For one, anthropic reasoning can be used 
to calculate a probability distribution for the observed values of some
fundamental property of the universe, such as the cosmological constant,
in models in which that property takes a variety of values with
varying probabilities (Efstathiou 1995; Vilenkin 1995; Weinberg 1996; 
Martel, Shapiro, \& Weinberg 1998, hereafter MSW).
Examples of such models
include those in which a state vector is derived for the universe which
is a superposition of terms with different values of the fundamental
property (e.g. Hawking 1983, 1984; Coleman 1988) and chaotic inflation
in which the observed big bang is just one of an infinite number of expanding
regions in each of which the fundamental property takes a different value
(Linde 1986, 1987, 1988). In models like these, the probability of
observing any particular value of the property is conditioned by the existence
of observers in those ``subuniverses'' in which the property takes that value.
This probability is proportional to the fraction of matter which is
destined to condense out of the background into mass concentrations
large enough to form observers -- i.e. the asymptotic collapsed fraction
for collapse into objects of this mass or greater.
MSW used this approach to offer a possible resolution of the infamous
``cosmological constant problem,'' one of the most serious crises of quantum
cosmology. Estimates of the size of a relic vacuum energy density $\rho_V$ from
quantum fluctuations in the early universe suggest a value which is many
orders of magnitude larger than the cosmic mass density today, and no
cancellation mechanism has yet been identified which would reduce
this to zero, let alone one so finely tuned as to leave the small but
{\it nonzero} value suggested by recent astronomical observations
(i.e. where the net $\rho_V$ is the sum of
a contribution from quantum fluctuations and a term 
$\Lambda/8\pi G$, where $\Lambda$ is the cosmological constant which appears
in Einstein's field equations) (Weinberg 1989; Carroll, Press, \& Turner 1992). 
MSW calculated the relative likelihood
of observing any given value of $\rho_V$ within the context of the
flat CDM model with nonzero cosmological constant, with the amplitude
and shape of the primordial power spectrum in accordance with current
data on the CMB anisotropy. Underlying this calculation was the notion
that values of $\rho_V$ which are large are unlikely to be observed since 
such values of $\rho_V$ tend to suppress gravitational instability and
prevent galaxy formation.
MSW found that a small, positive cosmological
constant in the range suggested by astronomical evidence is actually 
a reasonably
likely value to observe, even if the a priori probability distribution
that a given subuniverse has some value of the cosmological constant
does not favor such small values. Similar reasoning can, in principle, be 
used to assess the probability of our observing some range of values for other
properties of the universe, too, in the
absence of a theory which uniquely determines their values
(e.g.  the value of $\Omega_0$; Garriga, Tanaka, \& Vilenkin 1998). 
In such calculations, the asymptotic collapsed fraction is a 
fundamental ingredient.

Aside from its importance 
in anthropic probability calculations like these, in which one needs to
know the state of the universe in the infinite future, the asymptotic
collapsed fraction
is also relevant as an approximation to the {\it present} universe, for
the following reason. In an Einstein-de~Sitter universe, in which
there is no freeze-out, the asymptotic collapsed fraction is unity. In
any other unbound
universe, there will be a freeze-out at some epoch. If we live
in such a universe, the freeze-out epoch could be either in the future
or in the past. However, 
if recent attempts to reconcile a number of the observed properties of
our universe with theoretical models of the background universe and
of structure formation by invoking an unbound universe 
with $\Omo<1$ are correct, then the freeze-out epoch
is much more likely to be in the past.
If it were in the future, then the matter
density parameter today would still be close
to unity, e.g. $\Omo>0.9$ or 
$\Omo>0.99$.\footnote{This is a subjective notion,
since there is no precise definition of the freeze-out epoch.
For a flat, universe with positive cosmological constant, for
example, spherical density fluctuations must have fractional overdensity
$\delta=(\rho-\bar\rho)/\bar\rho\geq(729\rho_V/500\bar\rho)^{1/3}$
in order to undergo gravitational collapse, where $\rho_V$ is the vacuum energy
density and $\bar\rho$ is the mean matter density (Weinberg 1987). In this
case, once $\bar\rho$ drops to a value of the order of $\rho_V$ or less, only
density enhancements which are already nonlinear will remain gravitationally
bound. As such, the ``freeze-out'' epoch corresponds roughly to the time
when $\bar\rho\approx\rho_V$. Recent estimates from measurements of 
distant Type Ia SNe, however, suggest values which, if interpreted in terms
of this model, are closer to $\bar\rho\la\rho_V/2$ 
(Garnavich et al. 1998a; Perlmutter et al.
1998), so ``freeze-out'' began in the past for this model.}
If so, then the observable consequences of the eventual departure of
the background model from Einstein-de~Sitter would be largely in the future,
as well. As such, the strong motivation for considering models with
$\Omo<1$ in order to explain a number of the observed properties
of our universe as described above would vanish. In short, the current 
interest in a universe with $\Omo<1$ is consistent with a value of 
$\Omo$ small enough that the epoch of freeze-out is largely in the past.
In that case, the asymptotic collapsed fraction 
should be a good approximation to the {\it present} collapsed
fraction. This quantity is of interest, for
example, since, by combining
it with the observed luminosity density of the universe, we can get a handle
on the average mass-to-light ratio of the universe, and the amount
of dark matter. The complementary quantity, the {\it uncollapsed} fraction, 
is of interest, too, since it determines the amount of matter left
behind as the intergalactic medium, observable in absorption and by
its possible contributions to background radiation. 
A knowledge of the amount of matter left uncollapsed
is also necessary in order to interpret observations of gravitational lensing
of distant sources by large-scale structure. In addition,
as we shall see, the dependence of the asymptotic collapsed fraction
on the equation of state of the background universe will imply that
theoretical tools, such as the Press-Schechter approximation, 
require adjustment in order to take proper account of the effect of
``freeze-out'' on the rate of cosmic structure formation.

In this paper, we compute the asymptotic collapsed fraction for unbound
universes, using an analytical model involving spherical top-hat 
density perturbations surrounded by shells of compensating 
underdensity, applied statistically to the case of
Gaussian random noise density fluctuations, a model introduced by 
MSW for the particular case of a flat universe with a cosmological constant.
We consider a generic cosmological model with 2 components, 
a nonrelativistic component whose mean energy density varies as
$\bar\rho\propto a^{-3}$, where $a$ is the FRW scale factor, and a
uniform, nonclumping component whose energy density varies as
$\rho_{\rm X}\propto a^{-n}$, where $n$ is non-negative.
In terms of the equations of state for these two components,
we can write this as $p_i=w_i\rho_ic^2$, where $\rho_i$ and $p_i$ are
the mean energy density and pressure contributed by
component $i$. For the nonrelativistic matter component, $w=0$,
while for component X, $-1\leq w\leq0$ is the physically allowed range 
in models in which the universe had a big bang in its past and the energy of
component X was not more important in the past than that of matter,
which corresponds to $n=3(1+w)$ and the range $0\leq n\leq3$. The latter
condition is necessary in order to be consistent with observations
of cosmic structure and the CMB anisotropy today.
Special cases of this model include models with a cosmological constant
($n=0$), domain walls ($n=1$), infinite strings ($n=2$), massive
neutrinos ($n=3$), and radiation background ($n=4$)
(although, as explained above, we shall exclude values of
$n>3$ in our treatment here). This generic model, or similar ones, have 
been discussed previously by many authors (e.g. Fry 1985; 
Charlton \& Turner 1987; Silveira \& Waga 1994; Martel 1995; 
Dodelson, Gates, \& Turner 1996; Turner \& White 1997; Martel \& Shapiro 1998). 
Recently, such models have been referred to as ``quintessence'' models
(e.g. Caldwell, Dave, \& Steinhardt 1998) or as models involving 
``dark energy.''\footnote{We note that in some models, the X component is not
entirely nonclumping: For massive neutrinos, for example, the
assumption that the X-component is nonclumping is a very good
approximation only 
for fluctuations of wavelength smaller than the ``free-streaming,'' or
``damping,'' length of the neutrinos and for epochs such that longer wavelength
fluctuations are still in the linear amplitude phase.}
The Friedmann equation for this model is
\begin{equation}
\biggl({\dot a\over a}\biggr)^2=H_0^2
\Biggl[(1-\Omo-\Omx)\biggl({a\over a_0}\biggr)^{-2}
+\Omo\biggl({a\over a_0}\biggr)^{-3}
+\Omx\biggl({a\over a_0}\biggr)^{-n}\Biggr]\,,
\end{equation}

\noindent where H is the Hubble constant, $\Omega=\bar\rho/\rho_c$, 
$\Omega_{\rm X}=\rho_{\rm X}/\rho_c$, $\rho_c=3H^2/8\pi G$, and subscripts
zero indicate present values of time-varying quantities.

In \S2, we derive the conditions that the cosmological parameters must 
satisfy in order for the background universe to qualify as an eternal,
unbound universe. In \S3, we compute the critical density 
contrast $\delta_c$, defined as the minimum density contrast a spherical
perturbation must have in order to be bound. 
In \S4, we derive the asymptotic
collapse fraction $f_{c,\infty}$ in an unbound universe,
using the model introduced by MSW involving compensated spherical
top-hat density fluctuations. In \S5, we compute $f_{c,\infty}$ 
using the Press-Schechter approximation, instead.
In \S6, we compare the predictions of the two models.
As we shall see, this comparison points up a fundamental limitation to
the validity of the ad hoc, over-all correction factor of 2 by which 
the Press-Schechter integral over positive initial density fluctuations is
traditionally multiplied so as to
recover a total collapsed factor which takes account of the
accretion of mass initially in underdense regions.
In particular, we shall derive this factor of 2 for the
Einstein-de~Sitter case, but show that the same factor of 2 in the
Press-Schechter formula {\it overestimates} the asymptotic collapsed fraction
for an unbound universe. To illustrate the importance of these results for
currently viable models of cosmic structure formation, we apply our model
in \S6 to two examples of the Cold Dark Matter (CDM) model, with $\Omega_0=0.3$
and $H_0=70\rm\,km\,s^{-1}Mpc^{-1}$, the open, matter-dominated model and 
the flat model with cosmological constant.

\section{CRITERIA FOR AN UNBOUND UNIVERSE}

The Friedmann equation~(1) describes the time-evolution of the scale factor
$a(t)$. The solutions of this equation can be grouped into four categories,
according to their asymptotic behavior at late times.
If the derivative $\dot a$, which is initially positive, remains
positive at all times, never dropping to zero,
then the universe is unbound.\footnote{The asymptotic value of
$\dot a$ in the limit $a\rightarrow\infty$ can be either finite 
or infinite}
This is the case,
for instance, in a matter-dominated universe with $\Omo<1$.
If, instead, $\dot a$ drops to zero as $a\rightarrow\infty$, then the
universe is marginally bound. This is the case for the Einstein-de Sitter
universe ($\Omo=1$, $\Omx=0$). If $\dot a$ drops to zero at a {\it finite} 
value
$a=a_t$, then two situations can occur: If the second derivative $\ddot a$
is negative at $a=a_t$, the universe will turn back and recollapse. This is
the case for a matter-dominated universe with $\Omo>1$. However if both
$\dot a$ and
$\ddot a$ are zero at $a=a_t$, then the universe asymptotically approaches 
an equilibrium state with $a=a_t$ at late times. This is the case of the 
de~Sitter universe with a positive cosmological constant, which initially
expands, and asymptotically becomes an Einstein static universe.

To determine in which category a particular model falls, we need to study
the properties of the Friedmann equation~(1). For convenience, we 
rewrite this equation as 
\begin{equation}
g(y)\equiv H_0^2y\dot y^2=(1-\Omo-\Omx)y+\Omo+\Omx y^{3-n}\,,
\end{equation}

\noindent where $y\equiv a/a_0=1/(1+z)$. Only non-negative values of $g$
are physically allowed.
Since $y>0$ after the big bang,
the condition $g=0$ is equivalent to $\dot y=0$ (or $\dot a=0$).
The first term in the right-hand-side of
equation~(2) can be either positive or negative, while
the last two terms cannot be negative.\footnote{We are ignoring the 
possibility that $\Omx<0$, as would be the case, for instance, in a 
universe with a negative cosmological constant} 
If $n=2$ or $\Omx=0$, then
the quantity $\Omx$ cancels out in equation~(2). This merely illustrates
the fact that a universe with a uniform component whose density
varies as $a(t)^{-2}$ (cf. a universe with infinite strings)
behaves exactly like a matter-dominated universe. Such a universe is bound, 
marginally bound, or unbound if $\Omo>1$, $\Omo=1$, or $\Omo<1$,
respectively. The case in which $\Omx\neq0$ and $n=3$ is exactly the same as 
that with $\Omx=0$, except that $\Omo$ is everywhere replaced by
$\Omo+\Omx$. In that case, the universe is bound, marginally bound,
or unbound according to whether $\Omo+\Omx>1$, $\Omo+\Omx=0$, or
$\Omo+\Omx<1$, respectively.
Let us now consider cases with $\Omx\neq0$ for which $n\neq2$ and $n\neq3$.

For $1-\Omo-\Omx\geq0$, the universe cannot be bound.
Clearly, if $1-\Omo-\Omx>0$,
then $g(y)>0$ for all $y$, and the universe is unbound
for any value of $n$. If $1-\Omo-\Omx=0$,
then the last term in equation~(2) will eventually dominate
(since we assume $n<3$).
Two situations can then occur. If $n>2$,
then $g(y)$ grows more slowly than $y$,
implying that $\dot y^2=g(y)/H_0^2y$ decreases as $y$ increases, reaching zero
as $y\rightarrow\infty$. This is the case of a marginally bound universe.
If $n<2$, $\dot y^2$ will eventually increase with $y$. 
The universe is then unbound.

Let us now focus on the case $1-\Omo-\Omx<0$. If $n>2$, 
then at small $y$, $g(y)>0$, but as $y$ increases,
the first term in
equation~(2) will eventually dominate the other terms, giving $g(y)<0$.
There will therefore be a change of sign of $g(y)$ 
at some finite value $y=y_t$
where $g(y_t)=0$. That corresponds to a bound universe. 
This leaves the interesting case of a universe with $1-\Omo-\Omx<0$ and
$n<2$. Since the slope of $g(y)$ at early times for any $n<2$ is
$(1-\Omo-\Omx)<0$, while at late times it is $(3-n)\Omx y^{2-n}>0$, $g(y)$
has a minimum at some intermediate value of $y$. When $g(y)$ is
zero at that intermediate value, this corresponds to the case of a 
marginally bound universe.
The various possibilities for the cases with $1-\Omo-\Omx<0$ and $n<2$
are shown in Figure~1. The top curve shows a case
for which $g(y)>0$ for all $y$, that is, an unbound universe. The bottom 
curve shows a case for which $g(y)$ drops to zero at a finite value of $y$.
In this case, the universe turns back and recollapses. It is therefore 
bound.\footnote{At large $y$, the function $g(y)$ becomes positive again,
indicating that there are possible solutions for $y$ large. These are
``catenary universes,'' sometimes referred as ``no big bang solutions.''
In such models, the universe contracts from an infinite radius, turns
back, and reexpands forever. These solutions are not considered to be
physically interesting.} The transition between these two cases,
a marginally bound universe, is illustrated by the middle curve in
Figure~1, which is tangent to the $y$-axis. At $y=y_t$, both the
function $g(y)$ and its first derivative $dg/dy$ vanish. The condition
for having a marginally bound universe is, therefore, given by
the following simultaneous equations,
\begin{eqnarray}
&&(1-\Omo-\Omx)y_t+\Omo+\Omx y_t^{3-n}=0\,,\\
&&(1-\Omo-\Omx)+(3-n)\Omx y_t^{2-n}=0\,.
\end{eqnarray}

\begin{figure}
\vspace{9.5cm}
\includegraphics{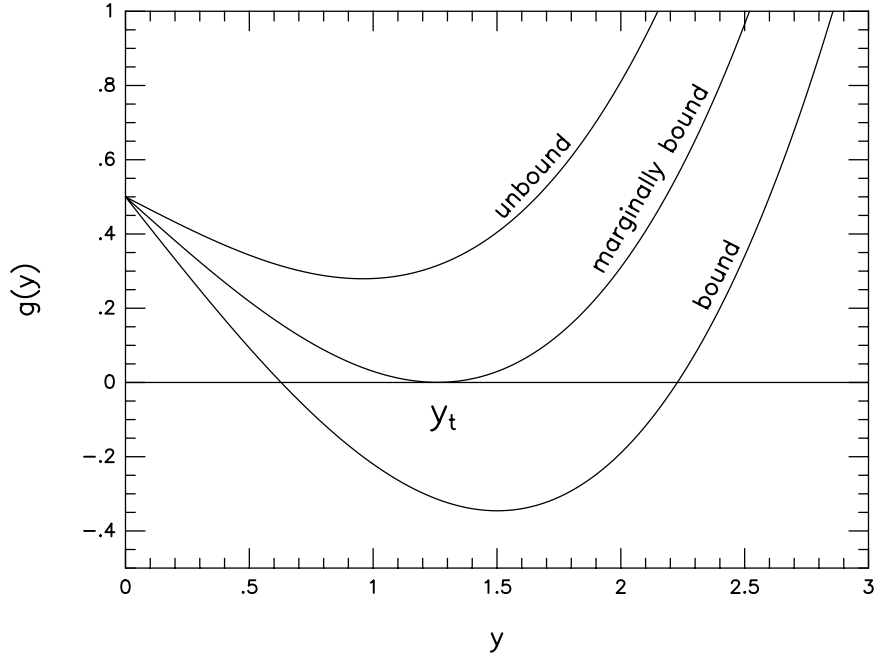}
\caption
{Schematic plot of function $g(y)$ 
versus $y$ for three different universes if $1-\Omo-\Omx<0$ and $n<2$: an
unbound universe, a marginally bound universe, and a bound universe.
The marginally bound case is characterized by the existence of
a point $y_t$ where $g=dg/dy=0$.}
\end{figure}

\noindent We can solve equation~(4) for $y_t$, and substitute 
this $y_t$ into equation~(3).
We get, after some algebra,
\begin{equation}
\biggl({\Omo+\Omx-1\over3-n}\biggr)^{3-n}
=\biggl({\Omo\over2-n}\biggr)^{2-n}\Omx\,.
\end{equation}

\noindent We can easily check some limiting cases. For a matter-dominated
universe ($\Omx=0$), equation~(5) gives $\Omo=1$ as the condition for
a marginally bound universe, as expected. For a universe with a nonzero
cosmological constant ($n=0$), equation~(5) reduces to
\begin{equation}
(\Omo+\lambda_0+1)^3={27\over4}\lambda_0\Omo^2\,,
\end{equation}

\noindent where we have replaced $\Omx$ by $\lambda_0$. This is actually
a well-known result (see, for instance, Glanfield 1966;
Felten \& Isaacman 1986; Martel 1990).

\section{THE CRITICAL DENSITY CONTRAST}

Consider, at some initial redshift $z_i\gg1$, a spherical perturbation
of density $\rho_i$=$\bar\rho_i(1+\delta_i)$ in an otherwise uniform
background of density $\rho_i$. 
Let us focus on positive density perturbations ($\delta_i>0$).
Clearly, if the background universe is bound or marginally bound, then
the perturbation is bound. However, if the background universe is 
unbound, then the perturbation can be either bound or unbound
depending upon the value of the initial density contrast $\delta_i$. 
Our goal in this section is to derive the {\it critical density contrast}
$\delta_{i,c}$, which is defined as the minimum value of $\delta_i$ for
which the perturbation is bound. To compute $\delta_{i,c}$, we make use
of the Birkhoff theorem, which implies that a uniform, spherically 
symmetric perturbation in an otherwise smooth Friedmann universe evolves
like a separate Friedmann universe with the same mean energy density
and equation of state as the perturbation.\footnote{Note:
For a {\it nonuniform} spherically symmetric perturbation,
every spherical mass shell evolves as it would in a universe with the same
mean energy density and equation of state as that of the average of the sphere 
bounded by that shell.} Pursuing this analogy, 
a perturbation with $\delta_i>\delta_{i,c}$ behaves like a bound universe,
a perturbation with $\delta_i<\delta_{i,c}$ behaves like an unbound universe,
and {\it a perturbation with $\delta_i=\delta_{i,c}$ behaves like a 
marginally bound universe}. We can then use the results of the
previous section to compute $\delta_{i,c}$.

First, we need to derive expressions for the ``effective cosmological
parameters'' of the perturbation. Notice first that an overdense perturbation
has been decelerating relative to the background between the big bang and
the initial redshift $z_i$. Hence, at $z=z_i$, the perturbation is expanding
with an ``effective Hubble constant'' $H_i'$ which is smaller than the
Hubble constant $H_i$ of the background universe. Assuming that the redshift
$z_i$ is small enough for linear theory to be accurate and for the universe to
resemble an Einstein-de Sitter universe ($\Omi\approx1$, $\Omxi\ll1$), but
late enough to allow us to neglect the linear decaying mode, we can easily
compute the relationship between $H_i'$ and $\delta_i$,
\begin{equation}
H_i'=H_i\biggl(1-{\delta_i\over3}\biggr)\,,
\end{equation}

\noindent (see, for instance, Lahav et al. 1991).
The effective density parameters of the perturbation are
then given by
\begin{eqnarray}
&&\Omi'={8\pi G\rho_i\over3{H_i'}^2}
={8\pi G\bar\rho_i(1+\delta_i)\over3H_i^2(1-\delta_i/3)^2}
={\Omi(1+\delta_i)\over(1-\delta_i/3)^2}\,,\\
&&\Omxi'={8\pi G\rho_{{\rm X}i}\over3{H_i'}^2}
={8\pi G\rho_{{\rm X}i}\over3H_i^2(1-\delta_i/3)^2}
={\Omxi\over(1-\delta_i/3)^2}\,.
\end{eqnarray}

Next, we need to find combinations of $\Omega_i'$ and $\Omega_{{\rm X}i}'$
that correspond to ``effective'' marginally bound universes.
For the cases for which $n<2$, this condition is given by equation~(5).
We now replace $\Omo$ and $\Omx$ by $\Omi'$ and $\Omxi'$ in 
equation~(5)\footnote{That equation was derived using the present values
of the density parameters, but it is of course valid at any epoch.} and
replace $\delta_i$ by $\delta_{i,c}$. This equation becomes
\begin{equation}
\biggl[{\Omi(1+\delta_{i,c})+\Omxi-(1-\delta_{i,c}/3)^2\over3-n}\biggr]^{3-n}
=\biggl[{\Omi(1+\delta_{i,c})\over2-n}\biggr]^{2-n}\Omxi\,.
\end{equation}

\noindent
Since $\delta_{i,c}\ll1$, we can expand this expression in powers 
of $\delta_{i,c}$ and keep only leading terms. We can then simplify this 
expression further by using the approximation $\Omi\approx1$.
Equation~(10) reduces to 
\begin{equation}
\biggl[{\Omi+\Omxi-1+5\delta_{i,c}/3\over3-n}\biggr]^{3-n}
={\Omxi\over(2-n)^{2-n}}\,.
\end{equation}

\noindent Notice that we had to keep the term $\Omi$ in the left hand side
because of the presence of the term $-1$, and that we cannot expand the 
left hand side in powers of $\delta_{i,c}$ because the quantity $\Omi+\Omxi-1$
might be as small as $\delta_{i,c}$. We now solve this equation for 
$\delta_{i,c}$, and get
\begin{equation}
\delta_{i,c}={3\over5}\biggl[
{(3-n)\Omxi^{1/(3-n)}\over(2-n)^{(2-n)/(3-n)}}+1-\Omi-\Omxi\biggr]\,.
\end{equation}

\noindent This gives the critical density contrast as a function of the
initial density parameters $\Omi$ and $\Omxi$. We can reexpress it as a
function of the present density parameters $\Omo$ and $\Omx$ and the
initial redshift, as follows: The initial density parameters are given by
$\Omi=8\pi G\bar\rho_i/3H_i^2=8\pi G\bar\rho_0(1+z_i)^3(H_0/H_i)^2/3H_0^2
=\Omo(1+z_i)^3(H_0/H_i)^2$ and $\Omxi=8\pi G\rho_{\rm X}/3H_i^2=
8\pi G\rho_{\rm X0}(1+z_i)^n(H_0/H_i)^2/3H_0^2=\Omx(1+z_i)^n(H_0/H_i)^2$. 
The ratio $(H_0/H_i)^2$
is given directly by equation~(1) (with $a_0/a_i=1+z_i$). We substitute these
expressions into equation~(12), and, using the fact that $z_i\gg1$, we
keep only the leading terms in $(1+z_i)^{-1}$. Equation~(12) reduces to
\begin{equation}
\delta_{i,c}={3\over5(1+z_i)}\Biggl[
{(3-n)\over(2-n)^{(2-n)/(3-n)}}\biggl({\Omx\over\Omo}\biggr)^{1/(3-n)}
+{1-\Omo-\Omx\over\Omo}\Biggr]\,.
\end{equation}

\noindent For the particular cases of a matter-dominated universe ($\Omx=0$)
or a flat universe with a nonzero cosmological constant 
($n=0$, $1-\Omo-\Omx=0$), we recover the results derived by
Weinberg (1987) and Martel (1994, eqs.~[7] and~[8]).

For cases in the range $2\leq n<3$, the condition for a marginally bound 
universe is $1-\Omo-\Omx=0$. We substitute equations~(8) and~(9) into this
expression, make the same approximations as above, and get
\begin{equation}
\delta_{i,c}={3\over5}(1-\Omega_i-\Omega_{{\rm X}i})\,.
\end{equation}

\noindent
In terms of the present density parameters and the initial redshift, this
expression reduces to
\begin{equation}
\delta_{i,c}={3(1-\Omo-\Omx)\over5\Omo(1+z_i)}\,.
\end{equation}

The case $n=3$ differs from all others in that the energy density of the X
component does {\it not} diminish relative to that of the ordinary matter
component as we go back in time. As such, we are never free to assume
that the early behavior of the top-hat is the same as it would be in the
absence of the X component. We shall, therefore, for simplicity, exclude
this case $n=3$ from further consideration here. 

\section{THE ASYMPTOTIC COLLAPSED FRACTION}

Our goal is to compute the fraction of the matter in the universe that
will eventually end up inside collapsed objects
(the asymptotic collapsed fraction). Clearly, this question only 
makes sense in unbound or marginally bound universes. In general, the
answer depends upon the mass scale of the collapsed objects being considered.
For cosmological models with Gaussian random noise initial conditions
(the usual assumption), the density contrast $\delta(\lambda)$ for
fluctuations of comoving length scale $\lambda$ is of order
$[k^3P(k)]^{1/2}$, where $k=2\pi/\lambda$ is the wavenumber, and
$P(k)$ is the power spectrum.
For a model such as Cold Dark Matter, for instance, the power spectrum
decreases more slowly than $k^{-3}$ at large $k$. Thus the density contrast
diverges at small scale. Normally, we eliminate small-scale perturbations
from the calculation by filtering the power spectrum at the mass scale of
interest, typically the mass required to form a galaxy. The density
fluctuations at that scale have a variance $\sigma^2$ given by
\begin{equation}
\sigma^2={1\over2\pi^2}\int_0^\infty P(k)\hat W^2(kR)k^2dk\,,
\end{equation}

\noindent where $\hat W$ is a window function, and $R$ is the comoving
radius of a sphere enclosing a mass in the unperturbed density field
which is equal to the mass scale of interest.
Assuming that the initial conditions are Gaussian, the fluctuation
distribution for {\it positive} values of $\delta$ is given by
\begin{equation}
{\cal N}(\delta)={2^{1/2}\over\pi^{1/2}\sigma}e^{-\delta^2/2\sigma^2}\,.
\end{equation}

\noindent Our problem consists of computing the asymptotic collapsed fraction
involving initially positive density fluctuations of mass equal to that
contained on average by a sphere of comoving radius $R$, together with
the additional mass which eventually accretes onto these positive density
fluctuations from initially underdense regions, starting
from initial conditions described by equations~(16) and~(17). In this
section, we consider the analytical model introduced by MSW. 
In the next section, we will consider
the well-known Press-Schechter approximation, instead.

Consider, at some early time $t_i$,
a spherical, top-hat matter-density fluctuation of volume $V$ and 
density contrast $\delta_i$, surrounded by a compensating shell of
volume $U$ and negative density contrast, such that the average
density contrast of the system top-hat + shell vanishes.
This model is parametrized by the shape parameter $s\equiv V/U$. 
If $\delta_i\geq\delta_{i,c}$, the top-hat core will collapse. Furthermore,
a fraction of the matter located outside the top-hat,
inside the shell, initially occupying
a volume $U'\leq U$, will be accreted by the top-hat. 
Since the density
is nearly uniform at early times, the asymptotic collapsed mass fraction
of this system is simply $(V+U')/(V+U)$. We now approximate the initial
conditions for the whole universe as an ensemble of these compensated 
top-hat perturbations, with a distribution of top-hat core
positive density fluctuations given by equation~(17), and we
neglect the interaction between perturbations. 
As discussed in MSW, the value of $s=0$ corresponds to the limit
in which each positive fluctuation is isolated,
surrounded by an infinite volume of compensating
underdensity (at a total density infinitessimally below the
mean value $\bar\rho$). For a flat universe with nonzero cosmological
constant, this case was treated by Weinberg (1996). The case $s=\infty$
corresponds to the limit of ``no infall'' in which
the additional mass associated with the compensating underdense volume
$V$ is negligible compared with that of the initial top-hat. This
case was considered for the flat universe with $\lambda_0\neq0$ by
Weinberg (1987). If $s=1$, however, the volume occupied by every positive
fluctuation is surrounded by an equal volume of compensating negative density
fluctuation. This is the case most relevant to the problem at hand, 
involving a Gaussian-random distribution of linear density fluctuations,
since the latter ensures that the volumes initially
occupied by positive and negative density fluctuations of equal amplitude 
are exactly equal. The full range of values of $s$, $0\leq s\leq\infty$,
was treated by MSW for the flat universe with $\lambda_0\neq0$,
with a special focus on $s=1$ as the case corresponding to Gaussian-random
noise initial conditions. The insensitivity
of the results for the anthropic probability calculations presented there
to the value assumed for $s$ suggests that the relative amount of
total collapsed fraction in universes with different values of $\rho_V$
may not be sensitive to the crudeness of the treatment of the
effect of one fluctuation on another. However, we 
will also present results here
for the full range of values of $s$, while noting that the value $s=1$ is
the most relevant to the case at hand of Gaussian random
density fluctuations.

Under these assumptions, the asymptotic collapsed fraction for the 
whole universe is given by
\begin{equation}
f_{c,\infty}={2^{1/2}s\over\pi^{1/2}\sigma_i}\int_{\delta_{i,c}}^\infty
{\delta e^{-\delta^2/2\sigma_i^2}d\delta\over\delta_{i,c}+s\delta}\,,
\end{equation}

\noindent 
where 
$\sigma_i$ is the value of $\sigma$ at time $t_i$ (MSW).
For bound and marginally bound universes (including, in 
particular, the Einstein-de~Sitter universe), $\delta_{i,c}=0$, 
and equation~(18) reduces trivially to $f_{c,\infty}=1$
for all values of $s$. Hence, the MSW model predicts that, in 
an Einstein-de~Sitter universe,
all the matter will eventually end up in collapsed objects. For unbound
universes, we change variables from $\delta$ to $x\equiv\delta^2/2\sigma_i^2$.
Equation~(18) reduces to
\begin{equation}
f_{c,\infty}={s\over\pi^{1/2}}\int_\beta^\infty
{e^{-x}dx\over sx^{1/2}+\beta^{1/2}}\,,
\end{equation}

\noindent where
\begin{equation}
\beta\equiv{\delta_{i,c}^2\over2\sigma_i^2}\,.
\end{equation}

\noindent This equation shows that the collapsed fraction $f_{c,\infty}$ is
unity only when $\beta=0$, which requires $\delta_{i,c}=0$.
However, in an unbound universe, 
$\delta_{i,c}$ is always positive. Hence, according
to the MSW model, the collapsed fraction in an unbound universe is always
less than unity. Notice that the dependence upon the cosmological 
parameters is entirely contained in the parameter $\beta$.
For any cosmological model, we can compute $\sigma_i$ using equation~(16)
and $\delta_{i,c}$ using either equation~(13) or ~(15). 
Since $\sigma_i\propto(1+z_i)^{-1}$
at large $z_i$ for any universe with $n<3$, the dependence on $z_i$
cancels out in the calculation of $\beta$, as it should: The asymptotic
collapsed fraction should not depend upon the initial epoch 
chosen for the calculation.

The size of the asymptotic collapse parameter $\beta$ determines not only
how large or small the collapsed fraction is but also how important
the increase of collapsed fraction is due to accretion from the surrounding
underdense regions. For small values of $\beta$, the asymptotic collapsed
fraction is close to unity because {\it both} the typical positive initial
density fluctuation {\it and} its fair share of the matter in surrounding 
regions of compensating underdensity collapse out {\it before} the effects 
of ``freeze-out''
suppress fluctuation growth. Hence, in this limit of small $\beta$, 
``freeze-out'' is unimportant and the results resemble that for an
Einstein-de~Sitter universe. For values of $\beta\ga1$, however,
the typical collapse occurs after ``freeze-out'' has begun to limit the
growth of density fluctuations. The large $\beta$ limit, in fact, is that in 
which only a rare, much-higher-than-average, positive density fluctuation is 
able to collapse out of the background before ``freeze-out'' prevents it,
and very little of the compensating underdense matter condenses out
along with it. For this large $\beta$ limit, equation (19) can be shown
to reduce to the following simple formula (see Appendix~A),
\begin{equation}
f_{c,\infty}(\beta\gg1)=\biggl({s\over s+1}\biggr)
{e^{-\beta}\over(\pi\beta)^{1/2}}\,.
\end{equation}

\section{THE ASYMPTOTIC LIMIT OF THE PRESS-SCHECHTER APPROXIMATION}

In the Press-Schechter approximation
(Press \& Schechter 1974; henceforth, ``PS''), the collapsed fraction at
time $t$ is estimated as follows: Consider a spherical 
top-hat perturbation with an initial linear
density contrast $\delta_i$ chosen such that this perturbation
collapses precisely at time $t$. The density contrast of that perturbation is
infinite at time $t$. 
However, if we estimate the density contrast at that epoch 
using linear perturbation theory, we obtain instead a finite 
value~$\delta=\Delta_c$, because linear theory underestimates the growth 
of positive fluctuations. The value of $\Delta_c$ is usually taken
to be $(3/5)(3\pi/2)^{2/3}=1.6865$, though this result is strictly correct
only for the Einstein-de~Sitter universe (cf. Shapiro, Martel, \& Iliev 1999,
and references therein).
A larger perturbation would collapse earlier, and linear theory would predict
that its density contrast exceeds $\Delta_c$ at time $t$. To compute the 
collapsed fraction at time $t$, we simply need to integrate over all
perturbations whose density contrast predicted by linear theory would exceed 
$\Delta_c$ at time $t$, using the distribution given by equation~(17). The
resulting expression, after multiplication by a factor of ``2'' to
correct for the fact that half the mass was initially in
underdense regions outside the positive density fluctuations, is
\begin{equation}
f_c^{\rm PS}={2^{1/2}\over\pi^{1/2}\sigma(t)}\int_{\Delta_c(t)}^\infty 
e^{-\delta^2/2\sigma(t)^2}d\delta\,.
\end{equation}

The introduction of this ad hoc correction factor of ``2'' in equation~(22)
is based on some assumption about the amount of matter located in 
unbound regions, either underdense or overdense, which is destined to be
{\it accreted} onto collapsed perturbations. Consider, for instance,
the case of an Einstein-de~Sitter universe
at late time. The critical density
contrast distinguishing a bound from an unbound density fluctuation
is zero, and therefore all overdense perturbations are bound and
will eventually collapse. Since, for Gaussian perturbations, the overdense
regions initially 
contain only half the mass of the universe, the asymptotic collapsed
fraction, without taking accretion into account, would be 
$f_{c,\infty}=1/2$. However,
it can easily be shown that in an Einstein-de~Sitter universe, all
matter in the universe will eventually end up inside bound objects.
Hence, for this particular case, the proper way to handle accretion is 
to multiply the collapsed fraction by a factor of 2. Equation~(22) is derived
by assuming that this factor of 2 is valid, not only for the asymptotic limit
of the Einstein-de~Sitter universe, but for all universes and at all epochs.
Hence, the PS approximation assumes that the total 
mass accreted by collapsed positive density fluctuations
is instantaneously equal to the total mass of these
collapsed objects themselves.

What is the asymptotic collapsed fraction according to this PS approximation?
We now change variables from $\delta$ to $x=\delta^2/2\sigma^2$.
Equation~(22) reduces to
\begin{equation}
f_c^{\rm PS}=
{1\over\pi^{1/2}}\int_{\beta_{\rm PS}}^\infty {e^{-x}dx\over x^{1/2}}\,,
\end{equation}

\noindent where
\begin{equation}
\beta_{\rm PS}(t)\equiv{\Delta_c(t)^2\over2\sigma(t)^2}\,.
\end{equation}

\noindent 
To compute the asymptotic collapsed fraction, $f_{c,\infty}^{\rm PS}$,
we need to take the limit of equations (23) and (24) as
$t\rightarrow\infty$. Consider a bound spherical perturbation, with the
values of its initial density contrast $\delta_i$ at initial time $t_i$ 
chosen so that it collapses at $t=\infty$. Call this value of $\delta_i$,
$\delta_{i,\infty}$. By definition, the quantity $\Delta_c$ at $t=\infty$ 
is given by
\begin{equation}
\Delta_c(\infty)=\delta_{i,\infty}{\delta_+(\infty)\over\delta_+(t_i)}\,,
\end{equation}

\noindent where $\delta_+(t)$ is the linear growing mode. Since this spherical
perturbation collapses at $t=\infty$, {\it the initial density contrast 
$\delta_{i,\infty}$ must be equal to the critical density contrast 
$\delta_{i,c}$}. If $\delta_{i,\infty} $ was less than $\delta_{i,c}$ the 
perturbation would not collapse at all, while if it was greater, 
the perturbation would collapse at a finite time. We can therefore 
replace $\delta_{i,\infty}$ by $\delta_{i,c}$ in equation~(25). Finally, 
we notice that the quantity $\sigma$ also evolves according to linear theory,
\begin{equation}
\sigma(\infty)=\sigma_i{\delta_+(\infty)\over\delta_+(t_i)}\,.
\end{equation}

\noindent Combining these results, we get
\begin{equation}
\beta_{\rm PS}(\infty)=
{\big[\delta_{i,c}\delta_+(\infty)/\delta_+(t_i)\big]^2
\over2\big[\sigma_i\delta_+(\infty)/\delta_+(t_i)\big]^2}
={\delta_{i,c}^2\over2\sigma_i^2}=\beta\,,
\end{equation}

\noindent 
(see eq.~[20]).
Hence, the PS $\beta_{\rm PS}$ parameter reduces
to the MSW $\beta$ parameter in the limit $t\rightarrow\infty$. Now, comparing
equations~(19) and (23), we see immediately that these equations are identical
in the limit $s\rightarrow\infty$. Notice that the product 
$\delta_{i,c}\delta_+(\infty)$ takes the undetermined form $0\cdot\infty$
in the case of an Einstein-de~Sitter universe. In this case, the
quantity $\delta_{i,c}\delta_+(\infty)/\delta_+(t_i)$ is equal to 
$(3/5)(3\pi/2)^{2/3}$, or 1.6865, at all times. In the Einstein-de~Sitter 
case, $\beta=\beta_{\rm PS}(\infty)=0$, and equations (19) and (23) are the 
same for all values of $s$; the asymptotic collapsed fractions in that case
are all equal to unity.

In the limit of large $\beta$, $f_{c,\infty}^{\rm PS}$ in equation (23), with
$\beta_{\rm PS}$ replaced by $\beta$, according to equation (27), can be shown
to reduce to the following simple formula (see Appendix A):
\begin{equation}
f_{c,\infty}^{\rm PS}(\beta\gg1)={e^{-\beta}\over(\pi\beta)^{1/2}}\,.
\end{equation}

\noindent A comparison of equations (21) and (28) reveals that the asymptotic
collapsed fraction $f_{c,\infty}^{\rm PS}$ according to the PS
approximation is just a factor of $(s+1)/s$ times $f_{c,\infty}$ according
to the MSW model, in the limit of large $\beta$.

\section{DISCUSSION AND CONCLUSION}

Our analytical result in equations (19) and (20) for the asymptotic collapsed
fraction in an eternal universe can be evaluated for any background
universe which satisfies the conditions given in \S2 which identify it
as an unbound universe. We need only specify the background universe and
the power spectrum of primordial density fluctuations, in order
to evaluate $\beta$. Before we
do this for a few illustrative cases, however, it is instructive to
evaluate the asymptotic collapsed fraction $f_{c,\infty}$ in general
as a function of $\beta$ and $s$, and compare $f_{c,\infty}$ to
the prediction of the PS approximation, $f_{c,\infty}^{\rm PS}$,
according to equations (23) and (27).

We have shown above that the asymptotic collapsed fraction predicted for an
eternal universe by the PS approximation differs from
that predicted here by the spherical model of MSW (as generalized to other
background universe cases) for $s=1$, the value of the shape parameter
appropriate for Gaussian random initial density fluctuations, with the 
exception of the Einstein-de~Sitter universe, for which 
$f_{c,\infty}^{\rm PS}=f_{c,\infty}=1$. For any eternal universe other than
Einstein-de~Sitter, in fact, the two approaches predict the same asymptotic
collapsed fraction only if $s=\infty$, instead. The fact that the two 
approaches generally predict different asymptotic collapsed fractions for
$s=1$ is not surprising, since the PS approximation never concerns itself with
the fraction of matter which is inside some gravitationally bound region
and is, hence, fated to collapse out, as the MSW model explicitly does. 
Instead, the PS approximation {\it assumes} that, as long as the matter is 
located within a region of average density which is high enough to make it
collapse according to the spherical top-hat model, it will not only collapse
but will also take with it an equal share of the matter outside this
region which was not initially overdense. This latter assumption is not
correct if the underdense matter is not all gravitationally bound to
some overdense matter. What is perhaps more surprising than this
disagreement between the two approaches for $s=1$ is the fact that
they {\it do} agree {\it for all models} if $s=\infty$.

The fact that in the limit $s\rightarrow\infty$ the MSW model
reduces to the asymptotic limit of the PS approximation is significant,
because the two models are based on different
assumptions. In the case of the PS approximation, a factor
of 2 is introduced to take accretion into account. In the MSW model, the
limit $s\rightarrow\infty$ corresponds to perturbations surrounded by 
underdense shells of negligible volume and mass. In this limit, there
is essentially no accretion. However, the volume filling factor of overdense
regions, which is 1/2 in the PS approximation, 
approaches unity in the limit $s\rightarrow\infty$
for the MSW model, resulting once again
in a factor of 2 in the expression for the collapsed fraction, but
for a different reason.

We have computed the collapsed fraction predicted by the MSW model
as a function of the parameter $\beta$, for various values of $s$,
by numerically evaluating equation~(19).
The results are plotted in Figure~2. In addition, the analytical expression
in equation (21) which is valid in the large $\beta$ limit is plotted
in Figure 2 for the case $s=1$. The analytical expression provides an
excellent fit to the exact results for the important case of $s=1$,
not only for large $\beta$, but for all $\beta\ga1$. The error even at $\beta=1$, for example, is only 15\%, while at $\beta=5$, the error is
reduced to 4.5\%. For comparison, we also show
the prediction of the PS approximation, according to 
equations (23) and (27). This curve is identical to the
curve for $f_{c,\infty}$ for the case $s=\infty$.
The point $\beta=0$, $f_{c,\infty}=1$ 
corresponds to the Einstein-de~Sitter universe.
As we see, all curves go through this point, indicating that the MSW
model predicts the correct asymptotic limit in this case, for any value
of $s$. The $s=1$ case is particularly important, since it is the only one
which gives equal filling factors to overdense and underdense perturbations,
a requirement for describing realistic Gaussian random initial conditions.

\begin{figure}
\vspace{9.5cm}
\includegraphics{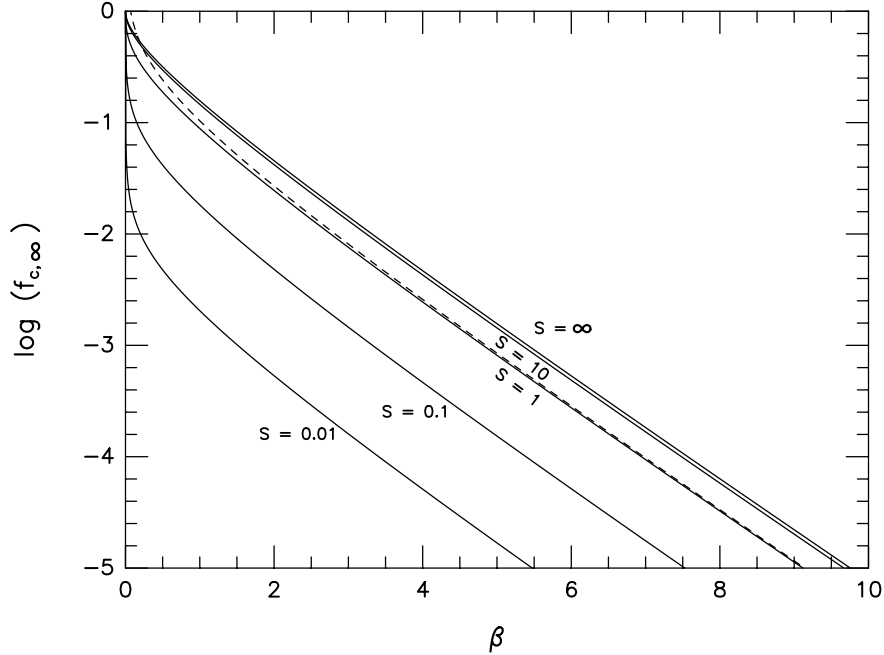}
\caption
{Asymptotic collapsed fraction $f_{c,\infty}$ versus $\beta$, calculated using
the MSW model for various values of the shape parameter $s$ (solid curves). 
The curve for $s=\infty$ is identical to $f_{c,\infty}^{\rm PS}$, the 
asymptotic limit ($t\rightarrow\infty$) of the Press-Schechter approximation.
Also plotted is the simple algebraic formula of the MSW model
in equation (21), derived
for the large $\beta$ limit, for the case $s=1$ (dashed curve).}
\end{figure}

Figure~2 shows that for finite values of $s$ and for $\beta>0$, the 
asymptotic collapsed fraction $f_{c,\infty}$ predicted by the MSW model is 
always less than the asymptotic collapsed fraction $f_{c,\infty}^{\rm PS}$ 
predicted by the PS approximation. For $s<1$,
this is not surprising, since in this limit the filling factor of the 
overdense regions is below the value of 1/2 assumed by the PS
approximation. However, if $s>1$, then the filling factor of overdense
regions exceeds 1/2, indicating that the bound
perturbations contain more mass in the MSW model than in the PS
approximation. In spite of this, we still have 
$f_{c,\infty}<f_{c,\infty}^{\rm PS}$. This
is caused by their different treatments of accretion. The MSW model
includes a detailed calculation of the amount of matter accreted by a
spherical top-hat, while the PS approximation simply assumes
that the accreted mass equals the initially overdense
mass, for all cosmological models.
Figure~2 suggests that this approximation can be quite crude in some
situations and greatly overestimate the amount of matter actually accreted.

To estimate this effect, we have computed, for the MSW model, the ``accretion
factor,'' $F_{\rm acc}$,
defined as the ratio of the total asymptotic collapsed fraction
$f_{c,\infty}$ divided by the asymptotic collapsed fraction 
$f_{c,\infty}^*$ that we
would obtain if accretion were neglected. (This factor $F_{\rm acc}$ is 2 for 
the PS approximation). We can easily compute $f_{c,\infty}^*$ by going 
back to the derivation of MSW and dropping the term in equation~(19) 
which represents the accreted matter. The resulting expression is
\begin{equation}
f_{c,\infty}^*(s,\beta)
={1\over\pi^{1/2}}\biggl({s\over s+1}\biggr)\int_\beta^\infty
{e^{-x}dx\over x^{1/2}}
=\biggl({s\over s+1}\biggr)f_{c,\infty}(s=\infty,\beta)\,.
\end{equation}

\noindent Hence, the accretion factor is
\begin{equation}
F_{\rm acc}(s,\beta)=\biggl({s+1\over s}\biggr)
{f_{c,\infty}(s,\beta)\over f_{c,\infty}(\infty,\beta)}\,.
\end{equation}

\noindent In the large $\beta$ limit, equations (21) and (30) indicate that
$F_{\rm acc}(s,\beta\gg1)=1$; in this limit, none of the matter in the
compensating underdense regions is able to condense out.

In Figure~3, we plot this accretion factor $F_{\rm acc}$
as a function of $\beta$, for various values of $s$. The factor
$F_{\rm acc}^{\rm PS}=2$ for the PS approximation is indicated
by the dashed line. For the MSW model,
the accretion factor depends mostly on the
amount of matter available in the shell surrounding the top-hat, which
goes to zero in the limit $s\rightarrow\infty$ and to infinity in
the limit $s\rightarrow0$. For the interesting case $s=1$ (underdense
and overdense regions with equal filling factors), we recover the 
PS limit $F_{\rm acc}=2$ at small $\beta$, but the value departs
rapidly from 2 at larger $\beta$. At $\beta=1$, for example,
the accretion factor drops 
to 1.125, indicating that the PS approximation overestimates
the amount of matter being accreted by a factor of 8!

\begin{figure}
\vspace{9.5cm}
\includegraphics{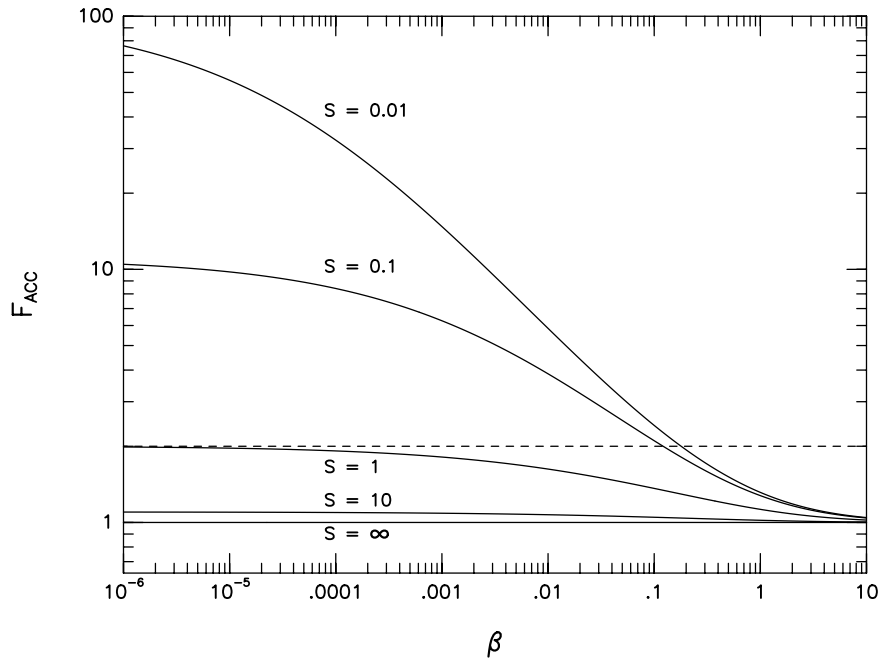}
\caption
{Accretion factor $F_{\rm acc}$ 
[$\equiv f_{c,\infty}\hbox{(accretion included)}/f_{c,\infty}\hbox{(no accretion)}$]
versus $\beta$ for the MSW model, for various values
of the parameter~$s$. The dashed line indicates the value of 2 which is used
in the Press-Schechter approximation.}
\end{figure}

To demonstrate the importance of this effect for actual cosmological models,
we consider two variations of the Cold Dark Matter (CDM) model:
(a) open, matter-dominated CDM ($\Omega_{\rm X0}=0$), and (b) flat
CDM with nonzero cosmological constant
($\Omega_{\rm X0}=\lambda_0=1-\Omega_0$,
$n=0$), both with an untilted primordial Harrison-Zel'dovich power 
spectrum\footnote{The exponent of the primordial power spectrum, which
is unity in the absence of tilt, is usually designated by
the letter $n$. It should not be confused by the exponent $n$ used
in this paper, which is introduced in equation~(1)}.
The primordial density fluctuation power spectrum for this model,
consistent with the standard inflationary cosmology and the measured
anisotropy of the cosmic microwave background according to
the {\it COBE} DMR experiment, is described 
in great detail in Bunn \& White (1997, and references 
therein). In the absence of tilt, this power spectrum 
(extrapolated to the present according to linear theory) is given by
\begin{equation}
P(k)=2\pi^2\biggl({c\over H_0}\biggr)^4\delta_H^2k^nT_{\rm CDM}^2(k)\,.
\end{equation}

\noindent where $c$ is the speed of light
and $T_{\rm CDM}$ is the transfer function, given by
\begin{equation}
T_{\rm CDM}(q)={\ln(1+2.34q)\over2.34q}\big[1+3.89q+(16.1q)^2+(5.46q)^3
+(6.71q)^4\big]^{-1/4}
\end{equation}

\noindent (Bardeen et al. 1986), with $q$ defined by
\begin{eqnarray}
q&=&\biggl({k\over{\rm Mpc}^{-1}}\biggr)\alpha^{-1/2}(\Omega_0h^2)^{-1}
\Theta_{2.7}^2\,,\\
\alpha&=&a_1^{-\Omega_b/\Omega_0}a_2^{-(\Omega_b/\Omega_0)^3}\,,\\
a_1&=&(46.9\Omega_0h^2)^{0.670}\big[1+(32.1\Omega_0h^2)^{-0.532}\big]\,,\\
a_2&=&(12.0\Omega_0h^2)^{0.424}\big[1+(45.0\Omega_0h^2)^{-0.582}\big]
\end{eqnarray}

\noindent (Hu \& Sugiyama 1996, eqs.~[D-28] and [E-12]), where 
$\Omega_b$ is the density parameter of the baryons, and
$\Theta_{2.7}$
is the temperature of the cosmic microwave background in units of 2.7K.
The quantity $\delta_H$ is given by
\begin{equation}
\delta_H=\cases{
1.95\times10^{-5}\Omega_0^{-0.35-0.19\ln\Omega_0}\,,& $\lambda_0=0$, no tilt;\cr
1.94\times10^{-5}\Omega_0^{-0.785-0.05\ln\Omega_0}\,,
& $\lambda_0=1-\Omega_0$, no tilt;\cr}
\end{equation}

Once the power spectrum is specified, we can compute the variance $\sigma^2$
of the present density contrast (i.e. as extrapolated to the present using
linear theory) as a function of the comoving length scale
or, equivalently, mass scale over which the density field is smoothed,
using equation (16). We then compute 
the parameter $\beta$ using equation~(20), where $\delta_{i,c}$ is given by
either equation~(13) or~(15), 
and $\sigma_i=\sigma\delta_+(z_i)/\delta_+(0)$. After 
some algebra, we get
\begin{equation}
\beta={9\over50\sigma^2\eta^2(\Omega_0,\lambda_0,z_i)}
\Biggl[3\biggl({\lambda_0\over4\Omega_0}\biggr)^{1/3}
+{1-\Omega_0-\lambda_0\over\Omega_0}\Biggr]^2\,,
\end{equation}

\noindent where the function $\eta(\Omega_0,\lambda_0,z)$ is defined
by
\begin{equation}
\eta(\Omega_0,\lambda_0,z)=(1+z){\delta_+(z)\over\delta_+(0)}
\end{equation}

\noindent (MSW). In the limit $z\gg1$, which we assume here,
the function $\eta$ becomes independent of $z$. For flat models 
($\Omega_0+\lambda_0=1$), MSW derived the following expression:
\begin{equation}
\eta(\Omega_0,1-\Omega_0,z\gg1)={6\lambda_0^{5/6}\over5\Omega_0^{1/3}}
\biggl[\int_0^{\lambda_0/\Omega_0}{dw\over w^{1/6}(1+w)^{3/2}}\biggr]^{-1}\,.
\end{equation}

\noindent For matter-dominated models, we can easily compute the function
$\eta$ using the expressions given in Peebles (1980). For open models, we get
\begin{equation}
\eta(\Omega_0,0,z\gg1)={2(1-\Omega_0)\over5\Omega_0}
\Biggl\{1+{3\Omega_0\over1-\Omega_0}+{3\Omega_0\over(1-\Omega_0)^{3/2}}
\ln\bigg[{1-(1-\Omega_0)^{1/2}\over\Omega_0^{1/2}}\Biggr]\Biggr\}^{-1}\,.
\end{equation}

The fraction of matter eventually collapsed into objects 
created by positive density fluctuations of mass greater than
or equal to some mass $M$ is entirely specified by the parameter $\beta$ 
evaluated for this mass scale as the density field filter mass.
Once $\beta$ is known, we can compute the asymptotic collapsed mass fractions
$f_{c,\infty}$ and $f_{c,\infty}^{\rm PS}$ 
using equations~(19) and (23), respectively.
We consider models with $H_0=70\,\rm km\,s^{-1}Mpc^{-1}$ ($h=0.7$), 
$\Omega_b=0.015h^{-2}$ (Copi, Schramm, \& Turner 1995), and $\Theta_{2.7}=1$.
We have computed $\sigma^2$ using equation~(16) with a top-hat window function,
\begin{equation}
\hat W(kR)={3\over(kR)^3}(\sin kR-kR\cos kR)\,.
\end{equation}

In Figure~4, we plot the variation of the asymptotic
collapse parameter $\beta$ with the filter mass $M$
[which corresponds to the length scale $R$ in equation~(42) according
to $M=4\pi R^3\rho_c\Omega_0/3$, or 
$M/M_\odot=1.163\times10^{12}R_{\rm Mpc}^3h^2\Omega_0$] for
two cases of interest: (a) open, matter-dominated, $\Omega_0=0.3$, and (b) flat
with cosmological constant, $\Omega_0=0.3=1-\lambda_0$. The value $\beta=1$
for these two cases corresponds to the mass scales 
$M/M_\odot=3.651\times10^{14}$ (open) and $5.778\times10^{14}$ 
(flat), respectively. For
$H_0=70\rm\,km\,s^{-1}Mpc^{-1}$ and density parameter $\Omega_0=0.3$ 
(assuming the shape parameter $s=1$, as required
for Gaussian random noise density fluctuations),
the open, matter-dominated CDM model and the flat CDM model with nonzero
cosmological constant yield mass fractions asymptotically collapsed into 
objects created by positive density fluctuations of mass greater than
or equal to the galaxy cluster mass-scale $10^{15}M_\odot$ of 0.0361 and
0.0562, respectively. These values of the asymptotic collapsed
fraction are only 55\% of the values
determined by the Press-Schechter approximation. These results have 
implications for the use of the latter approximation to compare the
observed space density of X-ray clusters today with that predicted
by cosmological models.

\begin{figure}
\vspace{9.5cm}
\includegraphics{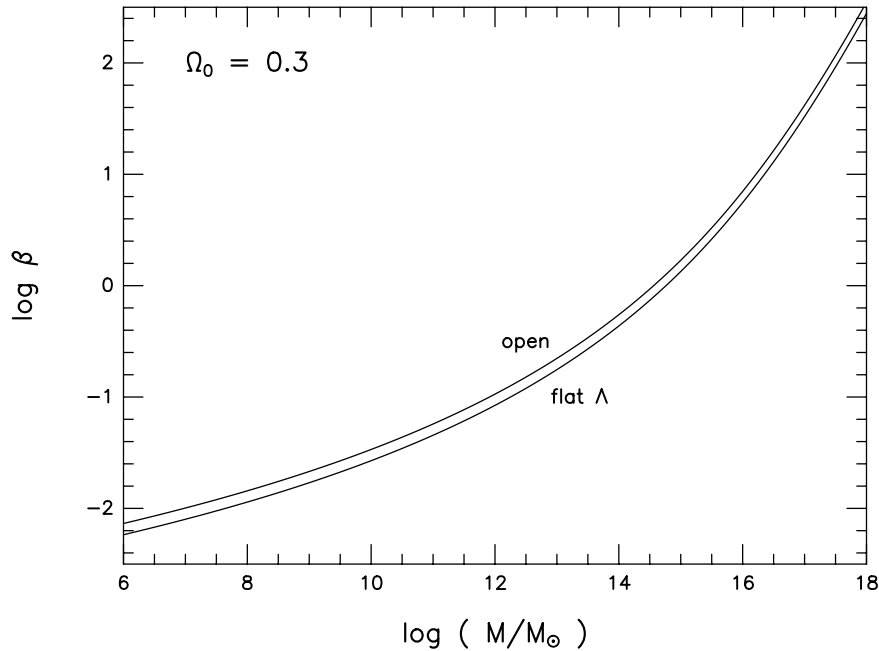}
\caption{Asymptotic collapse parameter $\beta$ versus the filter
mass scale $M$ for two {\it COBE}-normalized CDM models of interest
(with $\Omega_bh^2=0.015$, $H_0=70\,\rm km\,s^{-1}Mpc^{-1}$, and
a Harrison-Zel'dovich power spectrum): (a) open, matter-dominated 
($\Omega_0=0.3$, $\lambda_0=0$) and (b) flat, with cosmological constant
($\Omega_0=0.3$, $\lambda_0=0.7$).}
\end{figure}

We have also calculated the
asymptotic collapsed fractions $f_{c,\infty}$ and $f_{c,\infty}^{\rm PS}$
as a function of $\Omega_0$ (assuming $s=1$), for four different filter mass 
scales $M/M_\odot=10^6$, $10^9$, $10^{12}$, and $10^{15}$
(notice that the length scale
$R$ corresponding to a given mass scale varies with $\Omega_0$). The
results are shown in Figure 5. In Figure~5a, $f_{c,\infty}$ and
$f_{c,\infty}^{\rm PS}$ are each plotted separately, while in Figure~5b,
we plot the ratio $f_{c,\infty}^{\rm PS}/f_{c,\infty}$ to
demonstrate the extent to which the Press-Schechter approximation
overestimates the collapsed fraction, especially for cluster mass objects 
and above.

\begin{figure}
\vspace{12cm}
\includegraphics{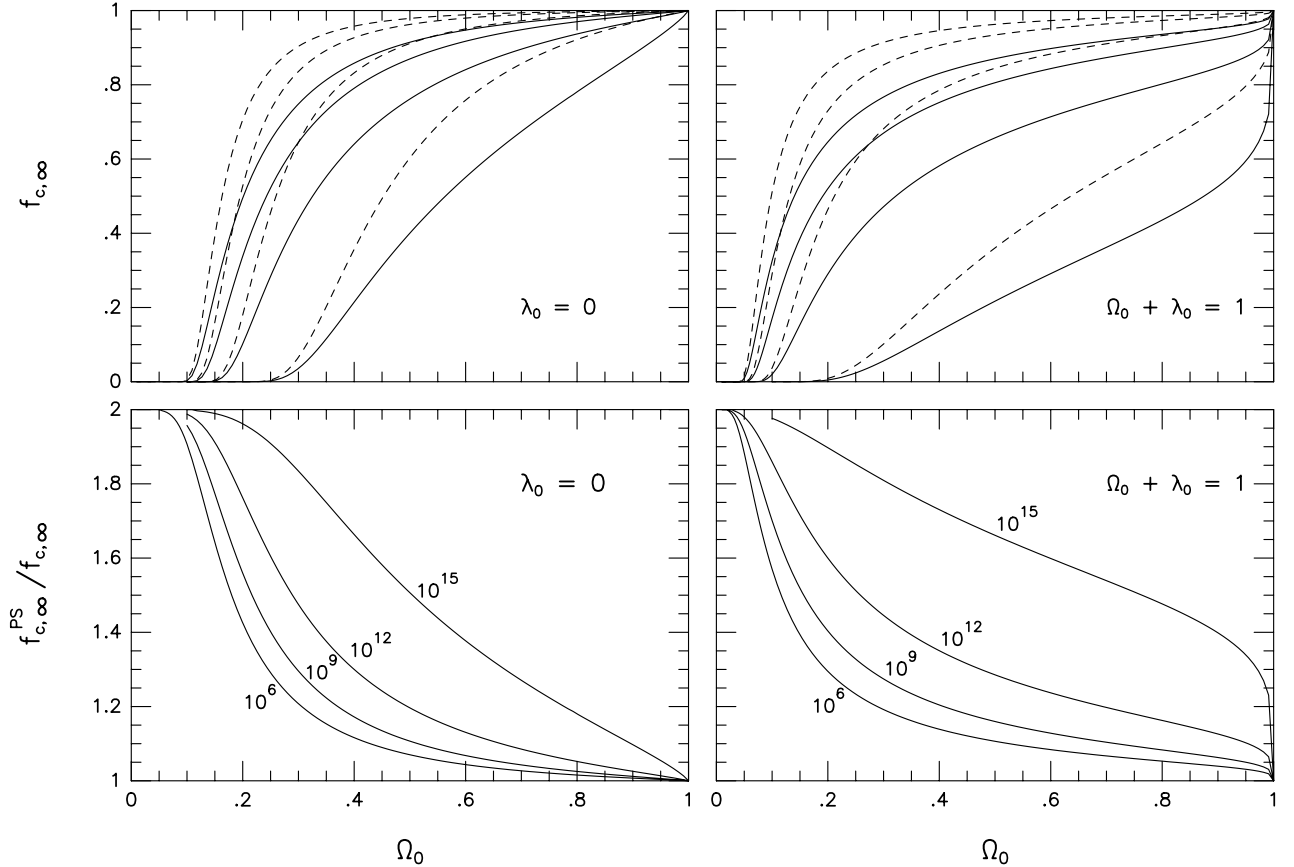}
\caption{(a) (top panels) Asymptotic collapsed fraction $f_{c,\infty}$ 
versus $\Omega_0$ for {\it COBE}-normalized CDM models
with a Harrison-Zel'dovich power spectrum, $H_0=70\,\rm km\,s^{-1}Mpc^{-1}$, 
and $\Omega_bh^2=0.015$: (i) open,
matter-dominated models (top panel), and (ii) flat models
with a nonzero cosmological constant
(bottom panel). The solid curves show the results obtained using
the MSW model. The dashed curves show the results obtained using the
Press-Schechter approximation. Each panel has four curves of each type,
corresponding to filter mass scales $M/M_\odot$ of $10^6$ (top curves),
$10^9$, $10^{12}$, and $10^{15}$ (bottom curves);
(b) (bottom panels) The ratio of the asymptotic collapsed fractions calculated
by the Press-Schechter approximation to those calculated using the
MSW model, for the cases shown in Fig.~5a. Curves are labelled with
the filter mass scales in solar mass units.}
\end{figure}

\section*{ACKNOWLEDGMENTS}
 
This work benefited greatly from our stimulating collaboration with
Steven Weinberg.
We are pleased to acknowledge the support of NASA Astrophysical Theory
Program Grants NAG5-2785, NAG5-7363, and NAG5-7812,
NSF Grant ASC~9504046, and a TICAM Fellowship in the summer of 
1998 for HM from the
Texas Institute of Computational and Applied Mathematics.

%

\appendix

\section{The Large $\beta$ Limit}

\subsection{The MSW Model}

The asymptotic collapsed fraction according to the MSW model is given by
\begin{equation}
f_{c,\infty}={s\over\pi^{1/2}}
\int_\beta^\infty {e^{-x}dx\over sx^{1/2}+\beta^{1/2}}\,.
\end{equation}

\noindent If we change variables using $x=\beta(1+w)$, then
equation~(A1) reduces to
\begin{equation}
f_{c,\infty}={s\beta^{1/2}e^{-\beta}\over\pi^{1/2}}
\int_0^\infty {e^{-\beta w}dw\over s(w+1)^{1/2}+1}\,.
\end{equation}

\noindent In the limit
$1/\beta\ll1$, we can always find a number $\alpha$ such that
$1/\beta\ll\alpha\ll1$. Since $\alpha\beta\gg1$, we can truncate the integral
in equation~(A2) at $w=\alpha$, because the 
exponential $e^{-\beta w}$ is negligible for larger values of $w$. Hence
\begin{equation}
f_{c,\infty}\approx{s\beta^{1/2}e^{-\beta}\over\pi^{1/2}}
\int_0^\alpha {e^{-\beta w}dw\over s(w+1)^{1/2}+1}\,.
\end{equation}

\noindent Since $\alpha\ll1$, the integration variable $w$ is always 
much smaller than unity, and we can replace $w+1$ by 1 in the denominator. 
The resulting integral yields
\begin{equation}
f_{c,\infty}\approx\biggl({s\over s+1}\biggr)
{e^{-\beta}(1-e^{-\beta\alpha})\over(\pi\beta)^{1/2}}\,.
\end{equation}

\noindent Since $\beta\alpha\gg1$, the term $e^{-\beta\alpha}$ is negligible.
The final expression is
\begin{equation}
f_{c,\infty}(\beta\gg1)\approx\biggl({s\over s+1}\biggr)
{e^{-\beta}\over(\pi\beta)^{1/2}}\,.
\end{equation}

\subsection{The PS Approximation}

The asymptotic collapsed fraction according to the PS approximation is given by
\begin{equation}
f_{c,\infty}^{\rm PS}={1\over\pi^{1/2}}
\int_\beta^\infty {e^{-x}dx\over x^{1/2}}\,.
\end{equation}

\noindent A change of variables to $w=x^{1/2}$ allows us to rewrite
equation (A6) as follows:
\begin{equation}
f_{c,\infty}^{\rm PS}={2\over\pi^{1/2}}
\int_{\beta^{1/2}}^\infty e^{-w^2}dw=1-{\rm erf}(\beta^{1/2})\,.
\end{equation}

\noindent For large $\beta$,
\begin{equation}
{\rm erf}(\beta^{1/2})\approx1-{e^{-\beta}\over(\pi\beta)^{1/2}}\,.
\end{equation}

\noindent Combining equations (A7) and (A8), we find
\begin{equation}
f_{c,\infty}^{\rm PS}(\beta\gg1)\approx{e^{-\beta}\over(\pi\beta)^{1/2}}\,.
\end{equation}


\begin{thebibliography}{}

\bibitem{}
Bahcall, N. 1999, in
Particle Physics and the Universe, in press (astro-ph/9901076)

\bibitem{}
Bardeen, J. M., Bond, J. R., Kaiser, N., \& Szalay, A. S. 1986, ApJ, 304, 15

\bibitem{}
Bunn, E. F., \& White, M. 1997, ApJ, 480, 6

\bibitem{}
Caldwell, R. R., Dave, R., \& Steinhardt, P. J. 1998, Phys.Rev.Lett.,
80, 1582

\bibitem{}
Carroll, S. M., Press, W. H., \& Turner, E. L. 1992, ARA\&A, 30, 499

\bibitem{}
Charlton, J. C., \& Turner, M. S. 1987, ApJ, 313, 495

\bibitem{}
Coleman, S. 1988, Nucl.Phys.B, 307, 867

\bibitem{}
Copi, C., Schramm, D. N., \& Turner, M. S. 1995, Science, 267, 192

\bibitem{}
Dodelson, S., Gates, E. I., \& Turner, M. S. 1996, Science, 274, 69

\bibitem{}
Efstathiou, G. 1995, MNRAS, 274, L73

\bibitem{}
Fry, J. N. 1985, Phys.Letters B, 158, 211

\bibitem{}
Felten, J. E., \& Isaacman, R. 1986, Rev.Mod.Phys., 58, 689

\bibitem{}
Garnavich, P. M. et al. 1998a, ApJ, 493, L53

\bibitem{}
Garnavich, P. M. et al. 1998b, ApJ, 509, 74

\bibitem{}
Garriga, J., Tanaka, T., \& Vilenkin, A., 1998, preprint (astro-ph/9803268)

\bibitem{}
Glanfield, J. R. 1966, MNRAS, 131, 271

\bibitem{}
Hawking, S. 1983, in Proc. 1983 Shelter Island Conf. on Quantum Field
Theory and the Fundamental Problems of Physics, ed. R. Jackiw et al.
(Cambridge: MIT Press)

\bibitem{}
Hawking, S. 1984, Phys.Lett.B, 175, 395

\bibitem{}
Hu, W., \& Sugiyama, N. 1996, ApJ, 471, 542

\bibitem{}
Krauss, L. 1998, preprint (astro-ph/9807376)

\bibitem{}
Lahav, O., Lilje, P. B., Primack, J. R., \& Rees, M. J. 1991, MNRAS, 251, 128

\bibitem{}
Linde, A. 1986, Phys.Lett.B, 175, 395

\bibitem{}
Linde, A. 1987, Phys.Scr., T15, 169

\bibitem{}
Linde, A. 1988, Phys.Lett.B., 202, 194

\bibitem{}
Martel, H. 1990, Ph.D. Thesis (Cornell University)

\bibitem{}
Martel, H. 1994, ApJ, 421, L67

\bibitem{}
Martel, H. 1999, ApJ, 445, 537

\bibitem{}
Martel, H., \& Shapiro, P. R. 1998, MNRAS, 297, 467

\bibitem{}
Martel, H., Shapiro, P. R., \& Weinberg, S. 1998, ApJ, 492, 29 (MSW)

\bibitem{}
Ostriker, J. P., \& Steinhardt, P. J. 1995, Nature, 377, 600

\bibitem{}
Peebles, P. J. E. 1980, The Large-Scale Structure of the Universe
(Princeton: Princeton University Press)

\bibitem{}
Press, W. H., \& Schechter, P. 1974, ApJ, 187, 425

\bibitem{}
Perlmutter, S. et al. 1999, ApJ, in press (astro-ph/9812133) 

\bibitem{}
Shapiro, P. R., Martel, H., \& Iliev, I. 1999, in preparation

\bibitem{}
Silveira, V., \& Waga, I. 1994, Phys.Rev.D, 50, 4890

\bibitem{}
Turner, M. S. 1999, in The Galactic Halo,
eds. B. K. Gibson, T. S. Axelrod, and M. E. Putman,
ASP Conference Series, Vol. 666, in press.

\bibitem{}
Turner, M. S., \& White, M. 1997, Phys.Rev.D, 56, R4439

\bibitem{}
Vilenkin, A. 1995, Phys.Rev.Lett., 74, 846

\bibitem{}
Weinberg, S. 1987, Phys.Rev.Lett., 59, 2067

\bibitem{}
Weinberg, S. 1989, Rev.Mod.Phys., 61, 1

\bibitem{}
Weinberg, S. 1996, in Critical Dialogues in Cosmology, ed. N. Turok
(Singapore: World Scientific), p. 1

\end{thebibliography}
\end{document}